\documentclass[12pt]{article}

\usepackage{amsmath}
\usepackage{amsfonts}

\usepackage{dsfont} 

\usepackage{mathtools} 

\usepackage{enumitem}

\usepackage{amsthm}
\newtheorem{theorem}{Theorem}
\newtheorem{lemma}[theorem]{Lemma}

\usepackage{hyperref}

\usepackage{multirow}
\usepackage{booktabs}
\usepackage{graphicx}
\usepackage{caption}
\usepackage{subcaption}
\usepackage[outdir=./]{epstopdf}
\usepackage{siunitx}
\usepackage[ruled,vlined]{algorithm2e}
\usepackage{colortbl}
\usepackage{tabulary}
\usepackage{etoolbox}
\newcommand{\expected}{\mathbb{E}} 
\newcommand{\prob}{\mathbb{P}} 

\newcommand{\timeValue}{k} 
\newcommand{\continuouTimeValue}{t} 

\newcommand{\graph}{G} 
\newcommand{\nodeSet}{V} 
\newcommand{\numNodes}{N} 
\newcommand{\node}{v} 
\newcommand{\edgeSet}{E} 
\newcommand{\degree}{d} 

\newcommand{\randNode}{X} 

\newcommand{\blueGroup}{\mathcal{B}} 
\newcommand{\numBlueNodes}{N^{\blueGroup}} 
\newcommand{\redGroup}{\mathcal{R}} 
\newcommand{\numRedNodes}{N^{\redGroup}} 

\newcommand{\partyFunction}{R} 
\newcommand{\nullHypothesis}{0} 
\newcommand{\alternativeHypothesis}{1} 
\newcommand{\beliefFunction}{H} 

\newcommand{\inAP}{\alpha} 
\newcommand{\outAP}{\beta} 

\newcommand{\redBirthProb}{r} 

\newcommand{\state}{\theta_{\timeValue}} 
\newcommand{\continuousTimeState}{\theta} 
\newcommand{\redState}{\theta^{\mathcal{R}}_{\timeValue}} 
\newcommand{\redContinuousTimeState}{\theta^{\mathcal{R}}}  
\newcommand{\blueState}{\theta^{\mathcal{B}}_{\timeValue}} 
\newcommand{\blueContinuousTimeState}{\theta^{\mathcal{B}}}  

\newcommand{\DotRedState}{\dot{\theta}^{\mathcal{R}}}
\newcommand{\DotBlueState}{\dot{\theta}^{\mathcal{B}}}

\newcommand{\probBlueZeroToOne}{p^{\mathcal{B}}_{\continuousTimeState}(0 \rightarrow 1)} 
\newcommand{\probBlueOneToZero}{p^{\mathcal{B}}_{\continuousTimeState}(1 \rightarrow 0)} 
\newcommand{\probRedZeroToOne}{p^{\mathcal{R}}_{\continuousTimeState}(0 \rightarrow 1)} 
\newcommand{\probRedOneToZero}{p^{\mathcal{R}}_{\continuousTimeState}(1 \rightarrow 0)} 

\newcommand{\homophily}{\rho}

\newcommand{\intertia}{\delta}

\usepackage{xcolor}

\usepackage{scicite}

\usepackage{times}

\newenvironment{sciabstract}{%
\begin{quote} \bf}
{\end{quote}}

\title{Dynamics of Affective Polarization: From Consensus to Partisan Divides}

\author
{Buddhika Nettasinghe,$^{1\ast}$ Allon G. Percus,$^{2}$ Kristina Lerman$^{3}$ \\
\\
\normalsize{\hspace{-1.2cm}$^{1}$University of Iowa,}\\
\normalsize{\hspace{-1.2cm}$^{2}$Claremont Graduate University}\\
\normalsize{\hspace{-1.2cm}$^{3}$USC Information Sciences Institute}\\
\\
\normalsize{\hspace{-1.2cm}$^\ast$To whom correspondence should be addressed; E-mail:  buddhika-nettasinghe@uiowa.edu.}
}

\date{}

\begin{document} 


\baselineskip24pt


\maketitle


\begin{sciabstract}

Politically divided societies are also often divided emotionally: people like and trust those with similar political views (in-group favoritism) while  disliking and distrusting those with different views (out-group animosity). This phenomenon, called affective polarization, influences individual decisions, including seemingly apolitical choices such as whether to wear a mask or what car to buy. We present a dynamical model of decision-making in an affectively polarized society, identifying three potential global outcomes separated by a sharp boundary in the parameter space: consensus, partisan polarization, and non-partisan polarization. 
Analysis reveals that larger out-group animosity compared to in-group favoritism,~i.e.~\emph{more hate than love}, is sufficient for polarization, while larger in-group favoritism compared to out-group animosity,~i.e.,~\emph{more love than hate}, is necessary for consensus.
We also show that, counter-intuitively, increasing cross-party connections facilitates polarization, and that by emphasizing partisan differences, mass media creates self-fulfilling prophecies that lead to polarization. Affective polarization also creates \emph{tipping points} in the opinion landscape where one group suddenly reverses their trends. Our findings aid in understanding and addressing the cascading effects of affective polarization, offering insights for strategies to mitigate polarization.

\end{sciabstract}

\section{Introduction}
American society has grown more ideologically divided, with Democrats and Republicans not only disagreeing on policy issues but also making dramatically different  choices about where to live and work, what products to buy, leisure activities to pursue~\cite{dellaposta2015liberals} or sports teams to support~\cite{swift}. 
Surveys also reveal a growing emotional divide, with members of each party increasingly disliking and distrusting the opposing party~\cite{iyengar2012affect,iyengar2015fear}. This phenomenon, called affective polarization, is manifested in people expressing warm feelings, i.e., \textit{in-group love}, towards their ideological allies  but negative feelings  and animosity, i.e., \textit{out-group hate}, to members of the opposing party. Over the last decade, cross-party antipathy has grown and now exceeds in-group love~\cite{druckman202218,finkel2020political}. 
The escalating partisan animosity poses a challenge to effective governing and the well-being of society. For example, during the COVID-19 pandemic individuals' trust and adherence to public health recommendations, like wearing a mask or getting vaccinated, were shaped by whether their own political party supported or opposed  those recommendations~\cite{grossman2020political}, hindering an effective response to the pandemic.

Research has shown that demographics alone cannot account for the partisan divide in beliefs and behaviors~\cite{iyengar2019origins,webster2017ideological,whitt2021tribalism}. 
Instead, these phenomena arise from collective social dynamics.
The tendency to associate with others who are similar, a process known as homophily, amplifies chance correlations between individual preferences and ideology, giving rise to a unified behavior within a group over time. 
This effect was used to explain the emergence of stereotypes like ``latte-drinking liberals'' and ``bird-hunting conservatives''~\cite{dellaposta2015liberals}. 
The rise of online media has further amplified social cleavages by enabling people to align their  information environments with their ideology. Similar to the mechanisms described above, these preferences tend to segregate people within ideologically-homogeneous communities, i.e., echo chambers~\cite{nikolov2015measuring,chen2021neutral}, which insulate them from opposing views and promote polarization. However, recent research has challenged this understanding~\cite{tornberg2022digital}, pointing to studies that show instead how increasing polarization can arise from exposure to opposing views.

This paper presents a model of information cascades in an affectively polarized social network composed of two groups (e.g., red and blue), where individuals within each group like and trust members of their own group (in-group love) and dislike and distrust members of the other group (out-group hate). 
When choosing between two possible choices~(e.g.,~wear a mask or not, get vaccinated or not, which team to support in the Superbowl), individuals observe their social connections 
and attempt to \textit{conform} to the choices of their in-group and \textit{oppose} choices made by members of their out-group. 
Depending on the size of the minority and majority groups, homophily~(preference of individuals to  connect to others of the same group), and the levels of in-group conformity and out-group opposition, several different long-term outcomes can emerge, marked by a sharp boundary:
global consensus (all individuals adopt the same choice), polarization~(party-line division of choices) 
and non-partisan polarization in which each group's choices are uniformly divided. We theoretically characterize the conditions under which such outcomes occur 
and provide numerical experiments that yield further insights.

Despite its simplicity, the model exhibits remarkably complex behaviors and reconciles seemingly contradictory findings from literature. The model explains how rapid collective transitions, or \emph{tipping points} in the opinion landscape~\cite{thurner2023new}, can emerge in social systems. It shows that opposition to the choices by members of the other party, driven by out-group hate, is a potent driver of polarization. When out-group hate is stronger than in-group love, no consensus is feasible. This may explain why disagreement on issues between Democrats and Republicans accelerated since 2012, when out-group hate exceeded in-group love in the U.S.~\cite{finkel2020political}. The model also explains why conventional wisdom-based approaches aimed at reducing polarization, such as connecting people from opposite parties, often backfire~\cite{bail2018exposure,tornberg2022digital}.  
{Specifically, our results corroborate the findings in \cite{siedlecki2016interplay} showing that consensus between two antagonistic communities can be achieved only when they are loosely connected. Beyond this, our analysis 
provides a comprehensive explanation for role of out-group hate, in-group love, cross-party connections and the initial beliefs in shaping opinions.}
Our work suggests that emphasizing partisan differences, even when they are small or non-existent, can fuel  polarization through a self-fulfilling prophecy. To counteract this, news media and social platforms could instead strive to diminish the perception of party-line differences  to impede actual polarization. For example, fostering connections between similar individuals from opposing parties may be one of the few effective methods to facilitate consensus. 

Our model is useful to understand the forms of divisions that emerge collectively from affective polarization, homophily and imbalanced party sizes and leads to new insights into polarization as well as methods to mitigate it. {The theoretical tractability of the model, which yields closed-form expressions for its dynamics, reduces the need to rely on large scale simulations to obtain such insights and may lead to new solutions to control polarization.}

\section{A Model of Information Cascades with Affective Polarization}
\label{sec:model_and_analysis}
We present a dynamical model of how people make choices in a social network~(e.g.,~to mask or support a sports team) by viewing the past choices of their in-group (e.g.,~members of their own party), which they approve of, as well as the choices of their out-group (e.g.,~cross-party members), which they oppose.
The choice dynamics lead to an information cascade which reaches a steady state of partisan polarization or consensus depending on group sizes and the levels of in-group love and out-group hate. 

Consider an undirected social network $\graph = (\nodeSet,\edgeSet)$ with $\numNodes = |V|$ individuals. Each individual (node) $\node \in \nodeSet$ has two binary attributes: a static binary attribute $\partyFunction(\node) \in \{0,1\}$ and a dynamic binary attribute $\beliefFunction_\timeValue(\node) \in \{\nullHypothesis,\alternativeHypothesis\}$ where $\timeValue$ denotes discrete-time. The static attribute represents the group (e.g., political) affiliation: $\node$ is red ($\node \in \redGroup$) if $\partyFunction(\node) = 1$; otherwise, $v$ is blue ($\node \in \blueGroup$). Let $\numBlueNodes = |\blueGroup|$ and $\numRedNodes = |\redGroup|$ denote the sizes of the two groups and $\redBirthProb = {\numRedNodes}/{\numNodes}$ denote the fraction of red nodes. 
The dynamic attribute $\beliefFunction_\timeValue(\node)\in\{\nullHypothesis,\alternativeHypothesis\}$ represents $\node$'s choice at time $\timeValue$~(e.g., wearing a mask vs not wearing a mask). 

At each time $\timeValue$ (where  $\timeValue= 0, 1, 2, \dots$), a node $\randNode_{\timeValue} \in \nodeSet$ chosen uniformly at random updates its choice by observing the 
choices of its neighbors.
Let
\begin{align}
    \begin{split}
    \label{eq:in_and_out_group_10_numbers}
    \degree_{\timeValue}^{in,\nullHypothesis}(\randNode_{\timeValue}) &= \sum_{(\randNode_{\timeValue},u) \in \edgeSet} \mathds{1}(\partyFunction(u) = \partyFunction(\randNode_{\timeValue}) \land \beliefFunction_{\timeValue}(u) = \nullHypothesis)/\degree(\randNode_{\timeValue})\\
    \degree_{\timeValue}^{in,\alternativeHypothesis}(\randNode_{\timeValue}) &= \sum_{(\randNode_{\timeValue},u) \in \edgeSet} \mathds{1}(\partyFunction(u) = \partyFunction(\randNode_{\timeValue}) \land \beliefFunction_{\timeValue}(u) = \alternativeHypothesis)/\degree(\randNode_{\timeValue})\\
    \degree_{\timeValue}^{out,\nullHypothesis}(\randNode_{\timeValue}) &= \sum_{(\randNode_{\timeValue},u) \in \edgeSet} \mathds{1}(\partyFunction(u) \neq \partyFunction(\randNode_{\timeValue}) \land \beliefFunction_{\timeValue}(u) = \nullHypothesis)/\degree(\randNode_{\timeValue})\\
    \degree_{\timeValue}^{out,\alternativeHypothesis}(\randNode_{\timeValue}) &= \sum_{(\randNode_{\timeValue},u) \in \edgeSet} \mathds{1}(\partyFunction(u) \neq \partyFunction(\randNode_{\timeValue}) \land \beliefFunction_{\timeValue}(u) = \alternativeHypothesis)/\degree(\randNode_{\timeValue})
    \end{split}
\end{align}
denote the number of in-group and out-group neighbors with choice-$0$ and choice-$1$ at time $\timeValue$ normalized by the total number of neighbors $\degree(\randNode_{\timeValue})$. Node $\randNode_{\timeValue}$ updates its choice at $\timeValue+1$ according to:
\begin{equation}
\label{eq:action}
    \beliefFunction_{\timeValue+1}(\randNode_{\timeValue}) = 
\begin{cases}
\nullHypothesis &\text{if } \inAP \left(\degree_{\timeValue}^{in,\alternativeHypothesis}(\randNode_{\timeValue}) - \degree_{\timeValue}^{in,\nullHypothesis}(\randNode_{\timeValue})\right) - \outAP\left(\degree_{\timeValue}^{out,\alternativeHypothesis}(\randNode_{\timeValue}) - \degree_{\timeValue}^{out,\nullHypothesis}(\randNode_{\timeValue})\right) < -\intertia \vspace{0.2cm}\\
\alternativeHypothesis &\text{if } \inAP \left(\degree_{\timeValue}^{in,\alternativeHypothesis}(\randNode_{\timeValue}) - \degree_{\timeValue}^{in,\nullHypothesis}(\randNode_{\timeValue})\right) - \outAP\left(\degree_{\timeValue}^{out,\alternativeHypothesis}(\randNode_{\timeValue}) - \degree_{\timeValue}^{out,\nullHypothesis}(\randNode_{\timeValue})\right) > \intertia\vspace{0.2cm}\\
\beliefFunction_{\timeValue}(\randNode_{\timeValue})  &\text{otherwise,} 
\end{cases}
\end{equation}
where $\inAP, \outAP, \intertia \in [0,1]$ are constant model parameters. Choices of all other nodes except $\randNode_{\timeValue} \in \nodeSet$ remain unchanged: for all $u \neq \randNode_{\timeValue}, \beliefFunction_{\timeValue+1}(u) = \beliefFunction_{\timeValue}(u)$.

The above stylized model aims to capture the dynamics of choices in an affectively polarized society. Consider a red node $\node$ deciding whether to wear a mask during the pandemic. The red neighbors~(in-group) that wear masks push $\node$ towards masking, whereas the red neighbors who do not wear masks push $\node$ towards not-masking. The out-group~(blue) neighbors have the opposite effect: blue masking neighbors push node $\node$ towards not-masking, whereas blue non-masking neighbors push the node towards masking. The relative strengths of these effects, \emph{in-group love} and \emph{out-group hate}, are quantified by $\inAP$ and $\outAP$, respectively. 
If the combined effect of out-group hate and in-group love exceeds $\intertia$ in favor of a certain choice~(1 or 0), then $\node$ adopts it. If not, it keeps it current choice. Thus, $\intertia$ quantifies the level of \emph{inertia} of a person, 
or the degree of social proof, including from the out-group, required to change the choice.
Also note from Eq.~\ref{eq:action} that, among the neighbors of $\node$ belonging to each group, only the difference between how many chose choice-0 and choice-1 matters and not the ratio. Even with the normalization in Eq.~\ref{eq:in_and_out_group_10_numbers},
50 out of a total of 100 masking blue neighbors will create a greater out-group effect for a red node than when one out of two blue neighbors masks. 

To analyze the dynamics, we examine the fraction of nodes in each group that have adopted choice-1 at time $\timeValue$. Formally, we define the state of the system at time $\timeValue$ as the column vector $\state = [\blueState, \redState]'$ where,
\begin{align}
    \begin{split}
    \label{eq:state}
    \blueState = \frac{\sum_{\node \in \nodeSet} \mathds{1}(\partyFunction(v) = 0 \land \beliefFunction_{\timeValue}(v) = \alternativeHypothesis)}{\sum_{\node \in \nodeSet} \mathds{1}(\partyFunction(v) = 0)}, \quad 
    \redState = \frac{\sum_{\node \in \nodeSet} \mathds{1}(\partyFunction(v) = 1 \land \beliefFunction_{\timeValue}(v) = \alternativeHypothesis)}{\sum_{\node \in \nodeSet} \mathds{1}(\partyFunction(v) = 1)}.
    \end{split}
\end{align}
\noindent 
Since the node $\randNode_\timeValue$ is chosen randomly at time $\timeValue$ to update its choice, the trajectory of the system $\state = [\blueState, \redState]', \timeValue = 0, 1, 2, \dots$ is also a random process. We show that the discrete-time stochastic trajectory $\state, \timeValue = 0, 1, 2, \dots$ can be approximated using the continuous-time deterministic trajectory of a differential equation under a few assumptions.
This differential equation representation of the stochastic model, called the \emph{limit mean differential equation} can thus be used to analyze the emergence of various patterns in the social network over sufficiently large time horizons. We will focus on two cases of practical interest: a fully connected network and a stochastic block model.

\subsection{Dynamics of the Model in a Fully Connected Network}
We first consider a fully connected social network $\graph = (\nodeSet, \edgeSet)$, where each node $v\in \nodeSet$ can observe the state of the system $\state = [\blueState, \redState]'$ at any time $\timeValue$. This occurs, for example, when people are informed about the prevalence of masking within each political party 
via daily news broadcasts   and make their  decisions to mask accordingly. 

In such a graph, 
the piece-wise interpolation\footnote{The piece-wise interpolation of $\continuousTimeState_\timeValue,  \timeValue = 0, 1, 2, \dots$ refers to the continuous time trajectory $\continuousTimeState^{\frac{1}{\numNodes}}(\continuouTimeValue) = \continuousTimeState_\timeValue$ for $\continuouTimeValue \in \left[\frac{k}{\numNodes}, \frac{k+1}{\numNodes}\right)$ for discrete time $\timeValue = 0,1,2,\dots$} of the discrete-time trajectory $\state = [\blueState, \redState]', \timeValue = 0, 1, 2, \dots$ can be approximated using the continuous-time trajectory $\continuousTimeState(\continuouTimeValue) = [\blueContinuousTimeState(\continuouTimeValue), \redContinuousTimeState(\continuouTimeValue)]', \continuouTimeValue\geq0$  of the following differential equation as the number of nodes in the graph $\numNodes$ is large:
\begin{align}
\label{eq:FC_ODE}
   &\begin{bmatrix}
\DotBlueState\\
\DotRedState
\end{bmatrix}   = 
\begin{bmatrix}
\left(1-\blueContinuousTimeState\right)\probBlueZeroToOne - \blueContinuousTimeState\probBlueOneToZero\\
\left(1-\redContinuousTimeState\right)\probRedZeroToOne - \redContinuousTimeState\probRedOneToZero\\
\end{bmatrix},
\end{align}
where,
\begin{align*}
&\probBlueZeroToOne = \mathds{1}\left(\inAP(1-\redBirthProb) \left(2\blueContinuousTimeState - 1\right) - \outAP\redBirthProb\left(2\redContinuousTimeState-1\right) > \intertia\right), \\
&\probBlueOneToZero = \mathds{1}\left(\inAP(1-\redBirthProb) \left(2\blueContinuousTimeState - 1\right) - \outAP\redBirthProb\left(2\redContinuousTimeState-1\right) < -\intertia\right), \\
&\probRedZeroToOne = \mathds{1}\left(\inAP\redBirthProb\left(2\redContinuousTimeState - 1\right) - \outAP(1-\redBirthProb)\left(2\blueContinuousTimeState-1\right) > \intertia\right), \\
&\probRedOneToZero = \mathds{1}\left(\inAP\redBirthProb\left(2\redContinuousTimeState - 1\right) - \outAP(1-\redBirthProb)\left(2\blueContinuousTimeState-1\right) < -\intertia\right).
\end{align*}


The intuition behind the differential equation in Eq.~\ref{eq:FC_ODE} is as follows. In a fully connected network, each node is a neighbor of all other nodes. Thus, the node-level statistics in Eq.~\ref{eq:in_and_out_group_10_numbers} can be written using the population statistics in Eq.~\ref{eq:state}. For a blue node $\randNode_\timeValue$, we can write   $\degree_{\timeValue}^{in,\alternativeHypothesis}(\randNode_{\timeValue}) = \blueState, \degree_{\timeValue}^{in,\nullHypothesis}(\randNode_{\timeValue}) = 1-\blueState, \degree_{\timeValue}^{out,\alternativeHypothesis}(\randNode_{\timeValue}) = \redState, \degree_{\timeValue}^{out,\nullHypothesis}(\randNode_{\timeValue}) = 1-\redState$. According to Eq.~\ref{eq:action}, a blue node $\randNode_\timeValue$ picks choice-1 when $\inAP(1-\redBirthProb) \left(2\blueState - 1\right) - \outAP\redBirthProb\left(2\redState-1\right) > \intertia$,~i.e.,~positive influence from the presence of choice-1 among in-group neighbors is larger than the negative influence from the presence of choice-1 among out-group neighbors by a margin of at least $\delta$. Similarly, a blue node picks choice-0 when $\inAP(1-\redBirthProb) \left(2\blueState - 1\right) - \outAP\redBirthProb\left(2\redState-1\right) < -\intertia$. Since a fraction $1-\blueState$ of blue nodes have choice-0 and a fraction $\blueState$ of blue nodes have choice-1, the expected rate of change of blue nodes with choice-1 $\blueState$ can thus be written as $\DotBlueState$ in Eq.~\ref{eq:FC_ODE}, and similarly for $\DotRedState$. When the network is large, 
the stochastic dynamics converge to the deterministic differential equation in Eq.~\ref{eq:FC_ODE} according to stochastic averaging theory. The formal proof of convergence
is given in Supplementary Information~(SI)~\ref{SI:proof_FC_Convergence}. Thus, for any initial state $\continuousTimeState(0) = [\blueContinuousTimeState(0), \redContinuousTimeState(0)]'$, the continuous-time trajectory $\continuousTimeState(\continuouTimeValue) = \continuousTimeState(0) + \int_0^\continuouTimeValue\dot{\continuousTimeState}(s)ds, \continuouTimeValue\geq0$ obtained using Eq.~\ref{eq:FC_ODE} approximates the stochastic model dynamics $\state = [\blueState, \redState]', \timeValue = 0, 1, 2, \dots$.


In the remainder of the paper, we rely on the 
differential equation in Eq.~\ref{eq:FC_ODE} and its generalizations to explore how polarized information cascades emerge in affectively polarized populations.

\subsection{Dynamics of the Model on a Social Network with Communities}
\label{subsec:SBM_Dynamics}
Next, we 
consider the case where the network $\graph = (\nodeSet, \edgeSet)$ is sampled from a stochastic block model with two communities. Specifically, each node is connected to a node in the same party with probability~$\homophily$ and a node in the other party with probability~$1-\homophily$, where $\homophily\in(0,1)$ is a constant model parameter. Thus, $\homophily$ quantifies the level of \emph{homophily}~\cite{mcpherson2001birds} of the individuals in the population: $\homophily>0.5$ implies that individuals are more likely to connect with others of the same party~(homophily), whereas $\homophily<0.5$ implies that individuals tend to mostly connect with members of the other party~(heterophily). When $\homophily=0.5$, the graph can be viewed as an Erd\H{o}s-R\'{e}nyi random graph with each edge being formed with a probability of $0.5$. 

Alternatively, $\homophily$ can be interpreted in the following way: each individual looks at a fraction $\homophily$ of their in-group members and a fraction $1-\homophily$ of their out-group members and makes a decision based on their choices. Thus, $\homophily$ might also be used to represent the balance of information an individual receives from the news media in terms of how well they represent the two parties: 
$\homophily>0.5$ means the news consumed by an individual over-represents views of the in-group~(relative to its size), while $\homophily<0.5$ means that the news over-represents the views of the out-group ~(relative to its size). When $\homophily = 0.5$, each group is represented in the news proportionate to its group size. 

The dynamics of the system $\state = [\blueState, \redState]', \timeValue = 0, 1, 2, \dots$ in a stochastic block model network can be approximated using the continuous-time trajectory of Eq.~\ref{eq:FC_ODE} with $\inAP$ replaced by $\inAP\homophily$ and $\outAP$ replaced by $\outAP(1-\homophily)$. In other words, the homophily $\homophily$ amplifies the effects of in-group love while reducing the effects of out-group hate. The exact differential equation for the stochastic block model is stated in SI~\ref{SI:SBM}.

\section{Results}

We analyze dynamics of the model and obtain insights about information cascades in an affectively polarized society. We first focus on a fully connected population with no inertia~(i.e.,~$\intertia = 0$) that starts from an initial state with no party-dependency~($\blueContinuousTimeState(0) = \redContinuousTimeState(0)$). 
The case $\intertia=0$ describes a highly reactive population where individuals choices are driven by the direction of the net effect of in-group neighbors and out-group animosity and not the amount. 
Then, we extend the results to more general settings with homophily, and party-dependent initial states~($\blueContinuousTimeState(0) \neq \redContinuousTimeState(0)$).

\subsection{Emergence of Polarization in a Fully Connected Network} 
\label{subsec:implications_of_AP}

Consider the case where choice-1 is initially equally popular in both groups~($\blueContinuousTimeState(0) = \redContinuousTimeState(0)$). 
This describes the early COVID-19 pandemic, when Democrats and Republicans were equally cautious about the disease 
and chose to mask. 
Remarkably, the long-term outcomes that emerge from a symmetric initial state can be characterized by just two quantities: the ratio of in-group love to out-group hate $\inAP/\outAP$ and the ratio of group sizes $\redBirthProb/(1-\redBirthProb)$.

\begin{figure}
\vspace{-1in}
    \centering
    \begin{subfigure}{\textwidth}
        \centering
        \includegraphics[width=\textwidth, trim=0.0in 0.1in 0.0in 0.0in, clip]{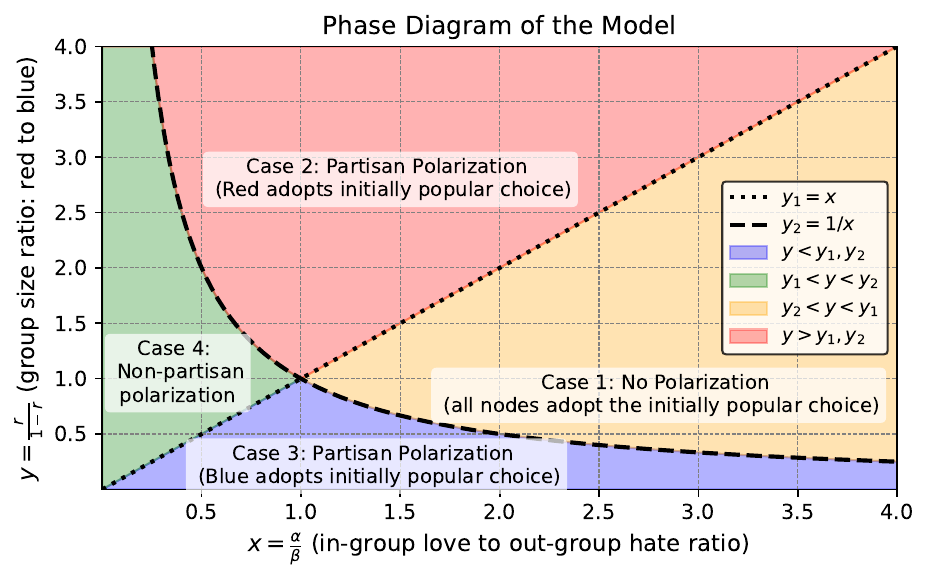}
    \end{subfigure}
    \begin{subfigure}{\textwidth}
        \centering
        \includegraphics[width=\textwidth, trim=0.0in 0.1in 0.0in 0.0in, clip]{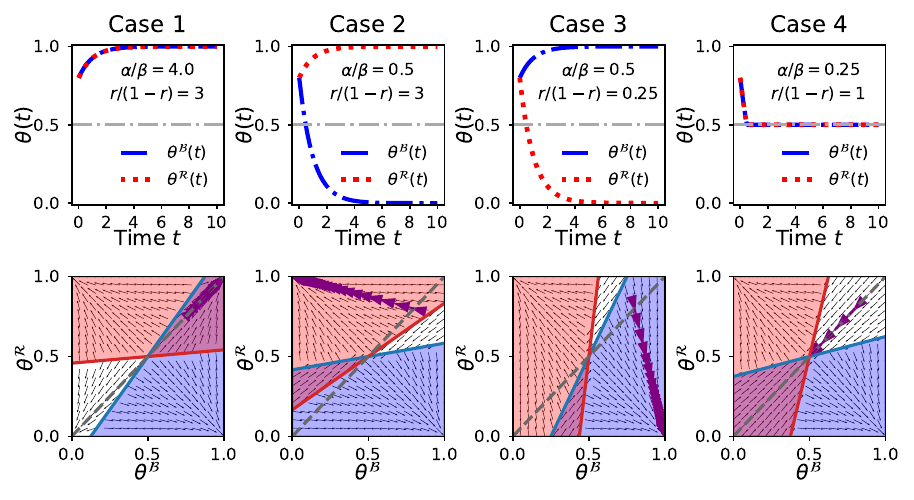}
    \end{subfigure}
    \caption{Phase diagram of the model~(top) and four example trajectories. The four different regions of the phase diagram (defined by the ratio of in-group love to out-group hate and the ratio of group sizes) lead to different 
    long-term outcomes
    in a fully connected network when both groups start from the same initial state ($\blueContinuousTimeState(0) = \redContinuousTimeState(0)$). The long-term outcomes are: {\bf (Case~1, yellow)} No Polarization, {\bf (Case~2, red / Case~3, blue)} Partisan Polarization, {\bf (Case~4, green)} Non-Partisan Polarization.
    Example trajectories in both time-domain and state space are shown below the phase diagram. The blue and red color areas in state space indicate regions  where $\blueContinuousTimeState(\continuouTimeValue), \redContinuousTimeState(\continuouTimeValue)$ increase~(i.e.,~regions where $\probBlueZeroToOne = 1$ and $\probRedZeroToOne=1$ according to Eq.~\ref{eq:FC_ODE}). The black arrows in state space plots indicate the path of the differential equation Eq.~\ref{eq:FC_ODE}. The purple arrows map the time domain trajectory to the state space.}
    \label{fig:Numerical_Results}
\end{figure}


\begin{theorem}[\emph{Information Cascades in a Fully Connected Network with Affective Polarization}]
\label{th:FC_Cases}
Consider Eq.~\ref{eq:FC_ODE} which represents the 
dynamics of the state of the population~$\continuousTimeState(\continuouTimeValue) = [\blueContinuousTimeState(\continuouTimeValue), \redContinuousTimeState(\continuouTimeValue)]'$ under the proposed model in a fully connected graph. Let $\intertia = 0$~(i.e.,~no inertia) and $\blueContinuousTimeState(0) = \redContinuousTimeState(0)$~(i.e.,~initial state is party independent). Then, the following statements characterize the asymptotic state of the system for various different values of $\inAP$~(level of in-group love), $\outAP$~(level of out-group hate) and $\redBirthProb$~(fraction of red nodes in the network):
\begin{itemize}
    \item {Case~1:} Let $\frac{\outAP}{\inAP}<\frac{\redBirthProb
    }{1-\redBirthProb} < \frac{\inAP}{\outAP}$. If $\blueContinuousTimeState(0) = \redContinuousTimeState(0)>0.5$, then $\lim_{\continuouTimeValue\longrightarrow\infty}\continuousTimeState(\continuouTimeValue) = [\blueContinuousTimeState_*, \redContinuousTimeState_*]' = [1,1]'$. If $\blueContinuousTimeState(0) = \redContinuousTimeState(0)<0.5$, then $\lim_{\continuouTimeValue\longrightarrow\infty}\continuousTimeState(\continuouTimeValue) = [\blueContinuousTimeState_*, \redContinuousTimeState_*]' = [0,0]'$~i.e.,~there is no polarization and both groups fully adopt the choice that was initially more popular.

    \item{Case~2:} Let $\frac{\redBirthProb
    }{1-\redBirthProb} > \frac{\inAP}{\outAP}$ and $\frac{\redBirthProb
    }{1-\redBirthProb} > \frac{\outAP}{\inAP}$. If $\blueContinuousTimeState(0) = \redContinuousTimeState(0)>0.5$, then $\lim_{\continuouTimeValue\longrightarrow\infty}\redContinuousTimeState(\continuouTimeValue) = [\blueContinuousTimeState_*, \redContinuousTimeState_*]' = [1,0]'$. If $\blueContinuousTimeState(0) = \redContinuousTimeState(0)<0.5$, then $\lim_{\continuouTimeValue\longrightarrow\infty}\continuousTimeState(\continuouTimeValue) = [\blueContinuousTimeState_*, \redContinuousTimeState_*]' = [0,1]'$~i.e.,~there is party-line polarization and the red-group~(which is the majority) fully adopt the choice that was initially popular while the blue-group fully adopt the other choice.

    \item{Case~3:} Let $\frac{\redBirthProb
    }{1-\redBirthProb} < \frac{\inAP}{\outAP}$ and $\frac{\redBirthProb
    }{1-\redBirthProb} < \frac{\outAP}{\inAP}$. If $\blueContinuousTimeState(0) = \redContinuousTimeState(0)>0.5$, then $\lim_{\continuouTimeValue\longrightarrow\infty}\redContinuousTimeState(\continuouTimeValue) = [\blueContinuousTimeState_*, \redContinuousTimeState_*]' = [0,1]'$. If $\blueContinuousTimeState(0) = \redContinuousTimeState(0)<0.5$, then $\lim_{\continuouTimeValue\longrightarrow\infty}\continuousTimeState(\continuouTimeValue) = [\blueContinuousTimeState_*, \redContinuousTimeState_*]' = [1,0]'$~i.e.,~there is party-line polarization and the blue-group~(which is the majority) fully adopt the choice that was initially popular while the red-group fully adopt the other choice.

    \item{Case~4:} Let $\frac{\outAP}{\inAP}>\frac{\redBirthProb
    }{1-\redBirthProb} > \frac{\inAP}{\outAP}$. If $\blueContinuousTimeState(0) = \redContinuousTimeState(0)>0.5$, then $\lim_{\continuouTimeValue\longrightarrow\infty}\continuousTimeState(\continuouTimeValue) = [\blueContinuousTimeState_*, \redContinuousTimeState_*]' = [0.5,0.5]'$. i.e.,~there is non-partisan polarization with half of each group adopting choice-1 and the remaining half adopting choice-0.
\end{itemize}
The limiting states in Cases~1-3~(consensus and polarization along party~lines) are locally asymptotically stable stationary states of the system in Eq.~\eqref{eq:FC_ODE} whereas the limiting state in Case~4 is an unstable stationary state of Eq.~\eqref{eq:FC_ODE}.
\end{theorem}



\subsubsection{Insights from Theorem~\ref{th:FC_Cases}} 
The four cases in Theorem~\ref{th:FC_Cases} shed light on the forms of polarization that can emerge in an emotionally divided population starting from a state with no group-level differences: (case~1) global consensus, where all nodes ultimately adopt the same choice, (case~2 and 3) party-line polarization, where the choices are split along party lines, and (case~4) non-partisan polarization, where each group is split evenly between the two choices. Below we consider additional  insights from Theorem~\ref{th:FC_Cases}. 

\noindent
\paragraph{Out-group hate is necessary for polarization:} 
Note from Fig.~\ref{fig:Numerical_Results}, that if $\outAP$ 
is approximately zero, then the network will always be in Case~1 which achieves consensus from any party-independent initial state $\blueContinuousTimeState(0) = \redContinuousTimeState(0) \neq 0.5$.

\noindent
\paragraph{Larger out-group hate relative to in-group love is sufficient for polarization:} 
When individual choices are driven more by a desire to oppose the out-group than a desire to conform to  the in-group, some form of polarization is unavoidable regardless of group sizes. As a result, in the region to the left  of the vertical line at $\inAP/\outAP=1$ in Fig.~\ref{fig:Numerical_Results}, consensus is not possible.
If out-group hate is very high compared to in-group love~($\inAP/\outAP \approx 0$ corresponding to case~4), then each group will be evenly split between the two choices. 
When the disparity between  $\inAP$ and $\outAP$ is not too large compared to group size disparity~(i.e.,~$\outAP/\inAP<\redBirthProb/(1-\redBirthProb)$ or $\inAP/\outAP>\redBirthProb/(1-\redBirthProb)$), polarization will emerge with the majority adopting the initially more popular choice and the minority  adopting the other choice~(Case~2 and Case~3 in Theorem~\ref{th:FC_Cases}).
Further, party-line polarization  is stable: a small deviation will push the system back to the polarized state as indicated by the arrows pointing to the polarized state in the state space plots of Fig.~\ref{fig:Numerical_Results}. Additional examples trajectories in the cases where polarization emerge are given in SI~Fig.~\ref{fig:LargeBeta}.


\noindent
\paragraph{Larger in-group love relative to out-group hate leads to consensus as long as the group imbalance is not too large:} When the two groups have the same size~(i.e.,~$\redBirthProb = 0.5$), Case~1 of Theorem~\ref{th:FC_Cases} shows that even a slightly larger in-group love compared to the out-group hate~(i.e.,~$\inAP>\outAP$) is sufficient for the network to adopt the initially popular choice, leading to consensus 
(see row~i of SI~Fig.~\ref{fig:LargeAlpha} for an example). 
Even with unequal group sizes, consensus can be achieved  with larger in-group love as long as the group imbalance is not large enough to push the system into Case~2 or Case~3. 
In other words, when $\inAP$ is sufficiently large compared to $\outAP$ , consensus can be achieved even when group sizes are not highly unequal 
(see row~ii of SI~Fig.~\ref{fig:LargeAlpha} for an example). Further, note that when $\outAP$ is negligible compared to $\inAP$, 
consensus is always achieved 
when both groups start from the same initial state (grey diagonal line in state space plots).
This highlights our  claim that out-group hate is crucial for any form of polarization to occur from a party independent initial state $\blueContinuousTimeState(0)=\redContinuousTimeState(0)$. However, even with high in-group love $\inAP>\outAP$, a large enough group imbalance ($\redBirthProb/(1-\redBirthProb)>\inAP/\outAP$ or $\redBirthProb/(1-\redBirthProb)<\outAP/\inAP$) can lead to polarization (as shown in row~iii of SI~Fig.~\ref{fig:LargeAlpha}). This observation emphasizes that \emph{more love than hate is necessary but not sufficient for consensus}.

\noindent
\paragraph{Majority cannot fully adopt the initially unpopular choice:} When $\redBirthProb>0.5$ (region above $y=1$ line in Fig.~\ref{fig:Numerical_Results}) and $\blueContinuousTimeState(0) = \redContinuousTimeState(0) > 0.5$ (i.e.,~choice-1 is initially more popular), there cannot be a case where all of the red-group adopts choice-0. In general, starting from a state $\blueContinuousTimeState(0) = \redContinuousTimeState(0)$ in a fully connected network, the majority cannot adopt the initially less popular choice.

\noindent
\paragraph{Small perturbations from non-partisan polarization (case~4) can lead to party-line polarization but not to consensus:} 
Consider Case~4 in Theorem~\ref{th:FC_Cases} where the population is evenly split between the two choices, regardless of group membership. This stationary state $\blueContinuousTimeState(\continuouTimeValue) = \redContinuousTimeState(\continuouTimeValue) = 0.5$ is unstable, and a small change in $\blueContinuousTimeState(\continuouTimeValue)$ or  $\redContinuousTimeState(\continuouTimeValue)$ can lead the population to polarize along party lines. 
This can be seen from state space plot corresponding to Case~4 in Fig.~\ref{fig:Numerical_Results}: a small deviation from $\blueContinuousTimeState(\continuouTimeValue) =  \redContinuousTimeState(\continuouTimeValue) = 0.5$ caused by a change of either $\blueContinuousTimeState(\continuouTimeValue)$ or $\redContinuousTimeState(\continuouTimeValue)$ will lead to party-line polarization. For example, if just a few red nodes switch to choice-1 from choice-0, $\blueContinuousTimeState(\continuouTimeValue)$ will converge to 1 and $\redContinuousTimeState(\continuouTimeValue)$ to 0.

Thus, even on a fully mixed population containing a majority and a minority that are not initially polarized, out-group hate and in-group love alone can lead to the emergence of a wide array of cascading choices. 

\subsection{Implications for Networks with Echo Chambers} 

Next, we consider the case where the network $\graph = (\nodeSet, \edgeSet)$ is sampled from a stochastic block model with two communities, where each node is connected to $\homophily$ fraction of their in-group members and $1-\homophily$ fraction of their out-group members, and $\homophily$ gives the homophily of the network.
Recall from Sec.~\ref{subsec:SBM_Dynamics} that the dynamics of the model with homophily can be obtained by replacing $\inAP$ and $\outAP$ in Eq.~\eqref{eq:FC_ODE} with $\inAP\homophily$ and $\outAP(1-\homophily)$, respectively. Consequently, replacing $\inAP$ and $\outAP$ in Theorem~\ref{th:FC_Cases} and Fig.~\ref{fig:Numerical_Results} with $\inAP\homophily$ and $\outAP(1-\homophily)$ leads to a characterization of the forms of polarization that can emerge in the presence of in-group love, out-group hate, homophily as well as a minority/majority division of the population. This is illustrated in SI~Fig.~\ref{fig:SBM_Regions}. We now discuss some insights on how these factors can collectively affect the emergence of polarization.

\noindent
\paragraph{Neutral homophily is indistinguishable from the fully-connected graph:} When people are neither homophilic nor heterophilic~($\homophily = 0.5$), the continuous-time trajectory in a stochastic block model is the same as the continuous-time trajectory in a fully connected graph given in Eq.~\eqref{eq:FC_ODE}~(since both sides of the inequalities inside indicator functions in Eq.~\eqref{eq:FC_ODE} would be multiplied by 0.5). Thus, Theorem~\ref{th:FC_Cases} 
as well as insights discussed in Sec.~\ref{subsec:implications_of_AP} are applicable not only to fully connected graphs but also to Erd\H{o}s-R\'{e}nyi random graphs where edges are formed in an independent and identically distributed manner.

\begin{figure}
\vspace{-1in}
    \centering
        \includegraphics[width=1\textwidth, trim=0.0in 0.12in 0.0in 0.0in, clip]{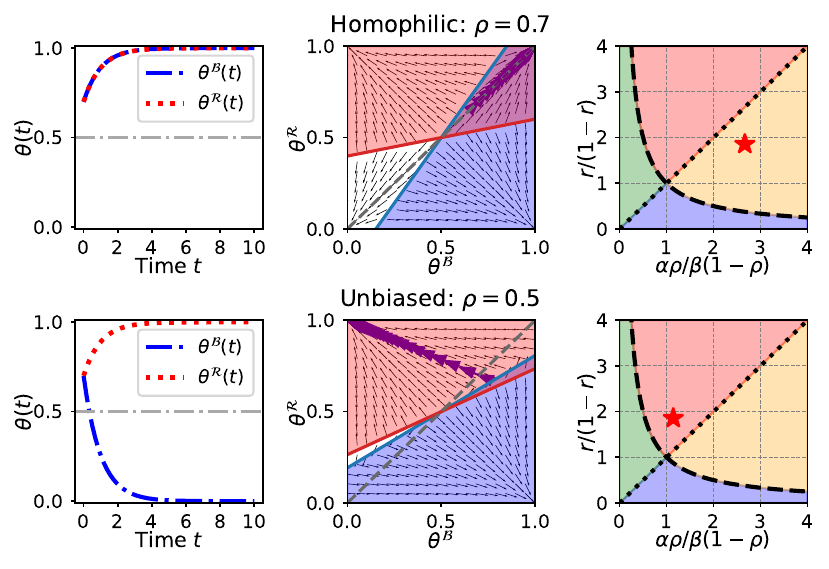}
    \caption{An illustration of how decreasing homophily can cause a party-line polarization. Both figures correspond to $\inAP = 0.8, \outAP = 0.7$~(larger in-group favoritism compared to out-group animosity) and $\redBirthProb = 0.65$~(a majority red group). First row corresponds to a homophilic network~(inter-group links are more likely to form than intra-group links) with $\homophily = 0.7$ whereas second row corresponds to an unbiased network~(all links are equally likely to form). Note that decreasing $\homophily$ from 0.7~(homophily) to 0.5~(unbiased) increases the effect of out-group hate and decreases the effect of in-group love on the choices, and pushes the social network from Case~1~(consensus) to Case~3~(party-line polarization) in Fig.~\ref{fig:Numerical_Results}~(with x-axis re-scaled as $\frac{\inAP\homophily}{\outAP(1-\homophily)}$).}
    \label{fig:IncreasingRho}
\end{figure}

\noindent
\paragraph{Highlighting the choices of the out-group in social networks may lead to polarization:} A typical approach to reducing partisan divisions calls for increasing the number of cross-party links. 
For example, consider the case where the two parties are approximately equal in size~($\redBirthProb \approx 0.5$) and $\inAP>\outAP$, which corresponds to Case~1 of Fig.~\ref{fig:Numerical_Results} where $\frac{\outAP}{\inAP}<\frac{\redBirthProb}{1-\redBirthProb} < \frac{\inAP}{\outAP}$. Thus, when an individual looks at the entire population (i.e.,~a fully connected graph) or an unbiased sample of the population~(i.e.,~an Erd\H{o}s-R\'{e}nyi random graph), universal consensus is achieved. Then, consider the case where the individual observes others in a biased manner, where each in-group member is observed with probability $\homophily$ and each out-group member with probability $1-\homophily$. If $\homophily<0.5$, the out-group will be over-represented compared to its size,
amplifying the effect of out-group hate while reducing the effect of in-group love. Thus, the population could move to the red~(Case~2) or blue regions~(Case~3) of Fig.~\ref{fig:Numerical_Results} where $\frac{\inAP\homophily}{\outAP(1-\homophily)}, \frac{\outAP(1-\homophily)}{\inAP\homophily}>\frac{\redBirthProb}{1-\redBirthProb}$ or $\frac{\inAP\homophily}{\outAP(1-\homophily)}, \frac{\outAP(1-\homophily)}{\inAP\homophily}<\frac{\redBirthProb}{1-\redBirthProb}$~i.e.,~partisan polarization can emerge starting from a uniform initial state where the choice is equally popular in both groups. Even a small increase in the number of cross-party links is likely to give rise to polarization~(Case 2 or Case 3) from a non-polarized state~(Case~1) when $\frac{\inAP\homophily}{\outAP(1-\homophily)} \approx \frac{\redBirthProb}{1-\redBirthProb}$ or $\frac{\outAP(1-\homophily)}{\inAP\homophily} \approx \frac{\redBirthProb}{1-\redBirthProb}$~(i.e.,~near the boundaries of Case~1 in the phase diagram of Fig.~\ref{fig:Numerical_Results} with x-axis re-scaled as $\frac{\inAP\homophily}{\outAP(1-\homophily)}$). Thus,\textit{ merely increasing the number of cross-party connections among the two groups may in fact facilitate polarization instead of consensus by amplifying the effect of out-group hate}. Figure~\ref{fig:IncreasingRho} shows two different trajectories of $\continuousTimeState(\continuouTimeValue)$ where the two groups start from the same initial state. Consensus is achieved for a homophilic network ($\homophily = 0.7$), where individuals get more information about the in-group, while polarization emerges in an unbiased network~($\homophily = 0.5$). This is because 
decreasing $\homophily$ from 0.7 to 0.5, pushes the network to Case~2 in 
Fig.~\ref{fig:Numerical_Results}~(with x-axis re-scaled as $\frac{\inAP\homophily}{\outAP(1-\homophily)}$).

In fact, increased exposure to the out-group~(i.e.,~decreasing $\homophily$) can bring divisions to a society already at global consensus. See SI~Fig.~\ref{fig:BreakingUnity} for an example. Note that global consensus remains at higher homophily (Case~1 in Fig.~\ref{fig:Numerical_Results}), 
and decreasing $\homophily$ to 0.5 makes the network unbiased but amplifies out-group hate, pushing it to Case~3, where the majority stays in the initial state but the minority adopts the choice that no one had chosen at the beginning. Further decreasing homophily makes the network highly heterophilic, where both groups focus largely on the out-group, pushing it to  Case~4. 
As 
this state is unstable, 
a small deviation causes polarization with one group adopting choice-1 and the other adopting choice-0. Thus, in a society with multiple ideologies, choices being driven by what the \emph{``opposition does"} more than what \emph{``our own group does''} can lead to divisive~(Case~2 and Case~3 in Fig.~\ref{fig:Numerical_Results}) and even unpredictable~(Case~4 in Fig.~\ref{fig:Numerical_Results}) division of choices for a society that was initially united. In practice, such situations occur when partisan information sources~(e.g.,~news organizations) emphasize the choices, decisions and actions of the out-group more than those of the in-group.

Relatedly, recall from Eq.~\ref{eq:FC_ODE} that when the two groups are approximately equal in size~(i.e.,~$\redBirthProb \approx 0.5$) and $\homophily = 0.5$~(unbiased network), people's choices are driven by $\continuousTimeState(\continuouTimeValue) = [\blueContinuousTimeState(\continuouTimeValue), \redContinuousTimeState(\continuouTimeValue)]'$~i.e.,~the prevalences of choice-1 in the in-group and out-group. If the popularity of choices is misrepresented in the information they receive at some time instant, that itself could lead to polarization. For example, consider \emph{latte drinking} as the choice and assume that it is equally prevalent among liberals and conservatives. However, if conservatives are selectively exposed to latte-drinking liberals, giving the perception that latte drinking is highly prevalent among them, that may cause them to give up lattes due to the out-group hate effect, and that in turn would lead liberals to further embrace it. Once this divergence takes off, it will be further amplified by the in-group love, leading to the eventual polarization of a seemingly non-partisan choice~\cite{dellaposta2015liberals}. Thus, even if a choice is not initially polarized, making it appear to be so in the news or on social media 
by selectively emphasizing the out-group,
can eventually lead to polarization in the form of a self-fulfilling prophecy. This serves as one possible explanation of why even traits that are historically non-partisan, such as the preferred choice of beverage, leisure activity, vocabulary, etc., can start to diverge along party lines when the prevalence of that trait in the opposite party is emphasized in the digital news~\cite{tornberg2022digital}. 

\subsection{Group-dependent Initial States}
\label{subsec:more_cases}

When choices are not initially identically distributed in the two groups
, several interesting phenomena can emerge. The differential equation in Eq.~\ref{eq:FC_ODE}~(and its generalization to stochastic block models) can be used to study such phenomena as well.
We begin by stating a result which characterizes conditions that lead to consensus from a party-dependent initial state.

\begin{theorem}[Consensus from Party-Dependent Initial States]
\label{th:consensus_conditions}
    Consider dynamics of the model on a fully connected graph given in Eq.~\ref{eq:FC_ODE} with $\intertia = 0$~(i.e.,~no inertia). Consensus emerges from a group-dependent initial state $\blueContinuousTimeState(0) \neq \redContinuousTimeState(0)$ if and only if,
    \begin{enumerate}
        \item  $\frac{\outAP}{\inAP}<\frac{\redBirthProb
    }{1-\redBirthProb} < \frac{\inAP}{\outAP}$, and,

        \item the initial state satisfies
        $\frac{\outAP\redBirthProb}{\inAP(1-\redBirthProb)}<\frac{2\blueContinuousTimeState(0) - 1}{2\redContinuousTimeState(0)-1} < \frac{\inAP\redBirthProb}{\outAP
(1-\redBirthProb)}$.
    \end{enumerate}
\end{theorem}

The first condition of Theorem~\ref{th:consensus_conditions} 
 states that the system has to be in Case~1 of Fig.~\ref{fig:Numerical_Results},
 which ensures that consensus is a stable steady state of the system. The second condition of Theorem~\ref{th:consensus_conditions} states that initial distribution of the choices within the groups cannot be too different from each other. The two conditions collectively ensure that consensus is reachable from the initial state. Any parameter configuration~($\inAP, \outAP, \redBirthProb$) or an initial state that does not satisfy the two conditions will give rise to polarization. The result further highlights the difficulties that lie in the path towards consensus in an affectively polarized society: even with high in-group love and balanced group sizes, the initial differences between the two parties can lead to polarized choices. In order to avoid this, social and news media through which people estimate the choice distributions must avoid emphasizing the differences between groups of different political ideologies.

\begin{figure}
\vspace{-1in}
    \centering
    \begin{subfigure}{\textwidth} 
        \includegraphics[width=\linewidth, trim=0.0in 0.1in 0.0in 0.0in, clip]{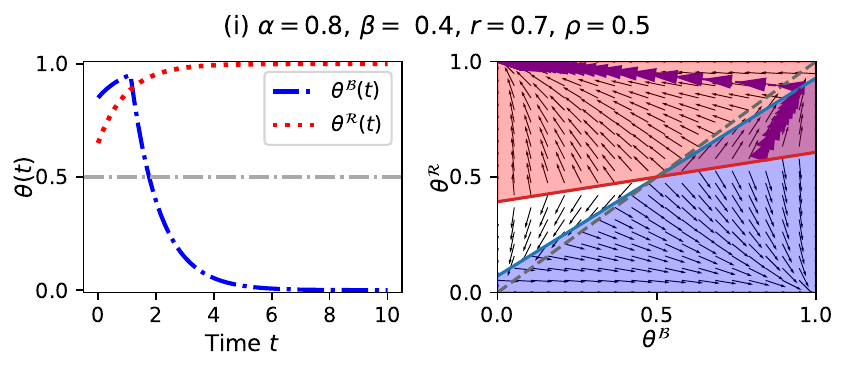}
    \end{subfigure}
    
    \begin{subfigure}{\textwidth} 
        \includegraphics[width=\linewidth, trim=0.0in 0.1in 0.0in 0.0in, clip]{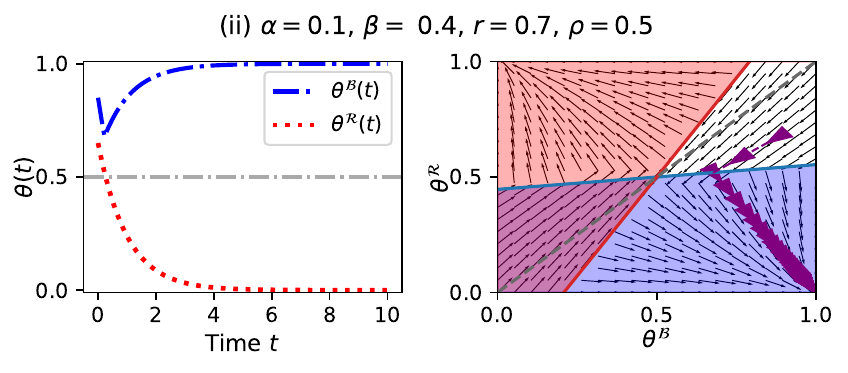}
    \end{subfigure}

    \begin{subfigure}{\textwidth} 
        \includegraphics[width=\linewidth, trim=0.0in 0.1in 0.0in 0.0in, clip]{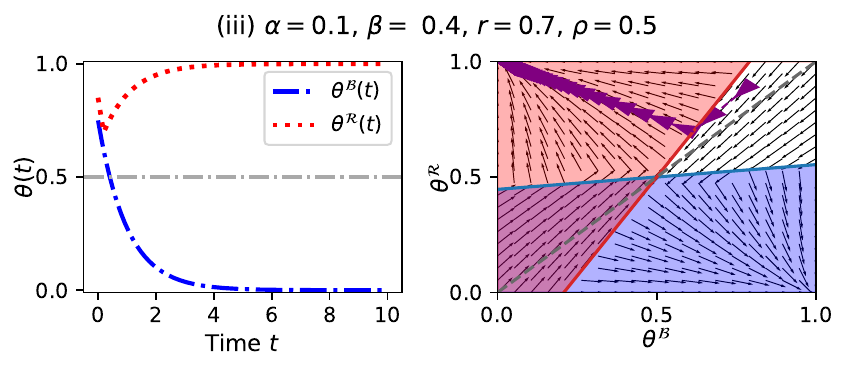}
    \end{subfigure}
    
    \caption{An illustration of three cases where the two groups start at different initial states~i.e.,~$\blueContinuousTimeState(0) \neq \redContinuousTimeState(0)$, and one group reverses its direction. In cases~i and ii, the minority blue group reverses its direction. In case~iii, the majority red group reverses its direction. The blue and red lines in state space indicate the \emph{tipping points} in opinion landscape where the respective group reverses its trend when the trajectory reaches it. The proposed model can demonstrate a variety of such phenomena when the initial states are different for the two groups.}
    \label{fig:DifferentInitialStates}
\end{figure}

\noindent
\paragraph{A group can flip:} When the groups start from different initial states, their trajectories can change direction. For example, consider the three cases in Fig.~\ref{fig:DifferentInitialStates}. 
In case~i
of Fig.~\ref{fig:DifferentInitialStates}, 
in-group love is higher than out-group hate~(i.e.,~$\inAP>\outAP$) and choice-1 is initially more prevalent within each group 
but to a different degree.
Due to higher in-group love, each group initially begins to embrace the choice-1 that is more popular within it. However, as this choice becomes more popular in the majority red group, the opposition  intensifies in the minority blue group, which starts to  adopt choice-0, leading to the eventual polarization. Interestingly, the \emph{flip} occurs when the population is very closer to consensus. This represents how political negotiations in an affectively polarized society can very unexpectedly break down even when they are on the verge of reaching bi-partisan agreements: the high presence of the same choice in both groups amplifies the effect of out-group hate. More precisely, in-group love is high enough to get closer to consensus~(due to the satisfied second condition of Theorem~\ref{th:consensus_conditions}), 
but it is not high enough to make consensus a stable stationary state~(due to violated first condition). More in-group love 
would drive both groups to consensus by focusing on unity within their own party rather than on hate towards the other party. Case~ii and case~iii of Fig.~\ref{fig:DifferentInitialStates} show scenarios with higher out-group hate where both conditions of Theorem~\ref{th:consensus_conditions} are violated. In case~ii, choice-1 is initially more prevalent in both groups but they both initially start adopting choice-0 due to higher out-group hate. However, as choice-0 becomes the more prevalent among the majority, the minority blue group starts adopting choice-1. Eventually, the trajectories converge in the opposite direction. Case~iii of Fig.~\ref{fig:DifferentInitialStates} shows a similar scenario where the majority red group reverses the trend. The theoretical tractability of the model Eq.~\ref{eq:FC_ODE} helps identify the exact trajectories for any initial state as seen from Fig.~\ref{fig:DifferentInitialStates}. 

\paragraph{The majority can eventually fully adopt the initially less popular choice:} 
Unlike the setting where both groups start in the same initial state, 
the majority can fully adopt the initially less popular choice when the two groups start in different initial states. For example, SI~Fig.~\ref{fig:DifferentInitialStates_majority} shows an example of a case where choice-1 is initially more popular among both groups:  
$\blueContinuousTimeState(0) = 0.9$ and 
$\redContinuousTimeState(0) = 0.6$. Also, 60\% of the nodes in the network are red, making it the majority. However, the red group eventually abandons choice-1 due to the out-group hate effect resulting from the high popularity of choice-1 among the blue group~(despite a smaller $\outAP$). In other words, due to high initial unity of the minority blue group, the majority red group is
driven more by a desire to oppose the blue party than to unite within their party. The minority blue group fully adopts choice-1 due to the higher in-group love effect created collectively by larger $\inAP$ and the high initial popularity of choice-1 within their group.

\section{Conclusion}





This paper introduced a dynamical model of decision making in a society where people trust the choices of those with same political views while distrusting the choices of those with opposing political views. 
The model is theoretically tractable and reveals the conditions for the emergence of consensus and partisan divisions from the initial state where there are no divisions. 
Our analysis highlights the importance of inter-group animosity in driving partisan division. 
Not only does out-group hate enable party-line polarization, but when it is larger than in-group love, consensus is no longer achievable.
{In particular, \emph{more hate than love is sufficient for partisan divisions while more love than hate is necessary for consensus}.}
When partisan mass media 
emphasize the choices of the out-group more than in-group~(i.e.,~
focusing on the other more than own group), it amplifies the effects of out-group hate and facilitates the emergence of polarization. This may create self-fulfilling prophesies where the perceptions of polarization actually give rise to polarization and explains why, counter to our intuition, cross-party exposure facilitates polarization rather than deterring it.
{High out-group hate can shatter consensus even when both parties are on the brink of agreement, a trend that is becoming increasingly common within emotionally polarized societies.}

The model and its theoretical tractability will also be useful to computational social scientists and network scientists to model the implications of affective polarization in future research and to gain insights on how to avoid its adverse implications on society.



\appendix
\renewcommand\thefigure{S\arabic{figure}}
\setcounter{figure}{0}

\section*{Supplementary Information}

\section{Proof of Convergence and Uniqueness}
\label{SI:proof_FC_Convergence}

\subsection{Outline of the Proof and Preliminaries}


\noindent
{\bf High-level idea of the proof:} The proof relies on the fact that the dynamics of $\state, \timeValue = 0,1,\dots$ are Markovian and the expected value of the next state given the previous state $\expected_\timeValue\left\{{\state}_{+1}\right\} = \expected\{{\state}_{+1} | \state\}, \timeValue = 0,1,\dots$ can be written as,
\begin{equation}
\label{eq:FC_Expected_value}
 \begin{bmatrix}
\expected_\timeValue\left\{{\blueState}_{+1}\right\}\\
\expected_\timeValue\left\{{\redState}_{+1}\right\}
\end{bmatrix}   = 
\begin{bmatrix}
{\blueState}\\
{\redState}
\end{bmatrix} + \frac{1}{\numNodes}\times\begin{bmatrix}
\left(1-\blueState\right)\probBlueZeroToOne - \blueState\probBlueOneToZero\\
\left(1-\redState\right)\probRedZeroToOne - \redState\probRedOneToZero\\
\end{bmatrix},
\end{equation}
where $\probBlueZeroToOne, \probBlueOneToZero, \probRedZeroToOne, \probRedOneToZero$ were defined in Eq.~\ref{eq:FC_ODE}. Therefore, the Markovian dynamics of the proposed model can be expressed as,
\begin{equation}
\label{SIeq:Overview}
{\state}_{+1} = {\state} + \frac{1}{N}(g(\state) + M_{k+1})
\end{equation}
where $$g(\state) = \begin{bmatrix}
g^{\mathcal{B}}(\state)\\
g^{\mathcal{R}}(\state)\\
\end{bmatrix} =\begin{bmatrix}
\left(1-\blueState\right)\probBlueZeroToOne - \blueState\probBlueOneToZero\\
\left(1-\redState\right)\probRedZeroToOne - \redState\probRedOneToZero\\
\end{bmatrix},$$
and ${M_k}$ is a martingale difference noise sequence. Eq.~\ref{SIeq:Overview} can be viewed as a stochastic approximation with constant step size $1/N$. 
Thus, for large $\numNodes$, the discrete time trajectory $\state, \timeValue = 0, 1, 2, \dots$ evolves without jumps and it converges
to the trajectory of the limit mean differential in Eq.~\ref{eq:FC_ODE}. For such constant step-size stochastic approximation algorithms, typical proof approach is to invoke a form of law of large-numbers and establish that the interpolated trajectory of Eq.~\ref{SIeq:Overview} converges weakly~(in distribution) to a differential equation of the form $\dot{\theta}(t) =  g(\theta(t))$ as the step size $1/N$ tends to 0. However, since $g(\cdot)$ is a discontinuous function, this typical approach that establishes (weak) convergence to an ordinary differential equation does not work. We establish the weak convergence of the interpolated stochastic trajectory of the model to a (deterministic) differential inclusion of the form $\dot{x}(\continuouTimeValue) \in  h(x(\continuouTimeValue))$ where $h(\cdot)$ is a set-valued map constructed using the discontinuous $g(\cdot)$. Any trajectory of the form $x(t) = x(0) + \int_0^t y(s)ds$ satisfying $y(t) \in h(x(t))$ for all $t$ is called a solution to the differential inclusion $\dot{x}(\continuouTimeValue) \in  h(x(\continuouTimeValue))$. Such solutions are called Filippov solutions to the discontinuous dynamical system $\dot{\theta}(t) =  g(\theta(t))$ (or Caratheodory solution of the differential inclusion $\dot{x}(\continuouTimeValue) \in  h(x(\continuouTimeValue))$)~\footnote{See \cite{cortes2008discontinuous} for a detailed introduction to discontinuous dynamical systems and their solution concepts.} We then show that due to the piece-wise continuous form of $g(\state)$, the solution is unique in all cases Except Case~4 of Theorem~\ref{th:FC_Cases}.

\noindent
{\bf Required results from literature:} The proof relies on two results from literature related to discontinuous dynamical systems that we state below. Let the distance between a continuous trajectory $z(\cdot)$ and the solution set $\mathcal{S}_T$ of a differential inclusion $\dot{x}(\continuouTimeValue) \in  h(x(\continuouTimeValue))$ be defined as, 
\begin{equation}
        l(z(\cdot),\mathcal{S}_T) \overset{\text{def}}{=} \underset{y(\cdot) \in \mathcal{S}_T}{\operatorname{inf}}     \underset{t \in [0,T]}{\operatorname{sup}}  ||z(t) - y(t)||.
\end{equation}

The following result from \cite{borkar2023stochastic} is used to establish the weak-convergence of the sample paths.
\begin{lemma}[\cite{borkar2023stochastic}[Adapted from Theorem~9.4]]
\label{Lemma:Borkar2023}
    Consider the stochastic approximation,
\begin{equation}
\label{SIeq:SA}
    x_{\timeValue+1} = x_\timeValue + a\left(g(x_\timeValue) + M_{\timeValue+1}\right), \timeValue \geq 0
\end{equation}
where $g(\cdot)$ is measurable and satisfies $||g(x)|| \leq C(1+||x||)$ for some $C>0$. Let 
\begin{equation}
    \label{SIeq:ConvexClosure}
    h(x) = \bigcap_{\epsilon>0}\bar{\text{co}}\left( g(y):||y-x||<\epsilon\right).
\end{equation}
where $\bar{\text{co}}$ denotes the convex closure.
Then, 
\begin{equation}
    l(x^{a}(\cdot)|_{[t',t'+T]}, \mathcal{S}_T) \overset{a \downarrow 0}{\longrightarrow} 0
\end{equation}
uniformly in $t'$ where $x^{a}(t)$ is the interpolated trajectory of the stochastic approximation algorithm and $\mathcal{S}_T$ is the solution set of the differential inclusion
\begin{equation}
    \dot{x}(\continuouTimeValue) \in  h(x(\continuouTimeValue)).
\end{equation}
\end{lemma}

We will also use \cite{cortes2008discontinuous}[Proposition~5] to establish the uniqueness of the solutions to the differential inclusion. At a high-level, \cite{cortes2008discontinuous}[Proposition~5] states that the Filippov solution of a piece-wise continuous differential equation with a discontinuous right-hand side~(i.e.,~the solutions to the differential inclusion constructed using that differential equation as in \ref{SIeq:ConvexClosure}) is unique if the trajectories that approach the boundary of a continuous region either slides along the boundary or cross into the next region.

\subsection{Proof of Convergence}
Consider the model proposed in Sec.~\ref{sec:model_and_analysis}. Note that the value of $\blueContinuousTimeState_{\timeValue+1} - \blueState$ can take three different values under three events:\\
Event~1: $\blueContinuousTimeState_{\timeValue+1} - \blueState = \frac{1}{\numBlueNodes}$ in the event that $\randNode_{\timeValue+1}$ is a blue node that takes action-0 at time $\timeValue$ and switches to action-1 at time $\timeValue+1$\\ Event~2: $\blueContinuousTimeState_{\timeValue+1} - \blueState = -\frac{1}{\numBlueNodes}$ in the event that $\randNode_{\timeValue+1}$ is a blue node that takes action-1 at time $\timeValue$ and switches to action-0 at time $\timeValue+1$\\Event~3:~$\blueContinuousTimeState_{\timeValue+1} - \blueState = 0$ in any event other than Event~1 and Event~3.

Let $\prob_{\timeValue}\{\cdot\}, \expected_{\timeValue}\{\cdot\}$ denote the probability measure and expected value conditional on all events that have occurred till time $\timeValue$. Consider the Event~1 first.   Note that the probability that $\randNode_{\timeValue}$ is a blue node with choice-0 at time $\timeValue$ is $\prob_{\timeValue}\left\{ \partyFunction(\randNode_{\timeValue+1})=0\land\beliefFunction_{\timeValue}(\randNode_{\timeValue+1}) = 0 \right\} = \frac{\numBlueNodes(1-\blueState)}{\numNodes}$. For a fully connected graph, note that the probability that a random blue node with choice-0 at time $\timeValue$ switches to the action choice-1 at time $\timeValue+1$ can be written as: 
\begin{align}
     &\prob_{\timeValue}\left\{\beliefFunction_{\timeValue+1}(\randNode_{\timeValue+1}) = 1|  \partyFunction(\randNode_{\timeValue+1})=0\land\beliefFunction_{\timeValue}(\randNode_{\timeValue+1}) = 0\right\}\\
     &=\prob_{\timeValue}\left\{\inAP \left(\degree_{\timeValue}^{in,\nullHypothesis}(\randNode_{\timeValue}) - \degree_{\timeValue}^{in,\alternativeHypothesis}(\randNode_{\timeValue})\right) - \outAP\left(\degree_{\timeValue}^{out,\nullHypothesis}(\randNode_{\timeValue}) - \degree_{\timeValue}^{out,\alternativeHypothesis}(\randNode_{\timeValue})\right) > 0|  \partyFunction(\randNode_{\timeValue})=0\land\beliefFunction_{\timeValue}(\randNode_{\timeValue}) = 0\right\}\\
     & = \mathds{1}\left(\inAP(1-\redBirthProb) \left(2\blueState - 1\right) - \outAP\redBirthProb\left(2\redState-1\right) > 0\right)\\
     &= \probBlueZeroToOne
\end{align}
Similarly, we also obtain,
\begin{align}
     &\probBlueOneToZero = \mathds{1}\left(\inAP(1-\redBirthProb) \left(2\blueState - 1\right) - \outAP\redBirthProb\left(2\redState-1\right) < 0\right).
\end{align}
Therefore, conditional on all events that have occurred till time $\timeValue$, the expected value of $\blueContinuousTimeState_{\timeValue+1}$ can be written as:
\begin{equation}
\expected_\timeValue\left\{{\blueState}_{+1} \right\}  =  {\blueState} +
\frac{1}{\numBlueNodes}\times\frac{\numBlueNodes\left(1-\blueState\right)}{\numNodes}\times
\probBlueZeroToOne - \frac{1}{\numBlueNodes}\times\frac{\numBlueNodes\blueState}{\numNodes}\times\probBlueOneToZero
\end{equation}

Following similar arguments for the red-group yields  similar expressions for $\expected_\timeValue\left\{{\redState}_{+1} \right\}$, which yields
\begin{equation}
 \begin{bmatrix}
\expected_\timeValue\left\{{\blueState}_{+1}\right\}\\
\expected_\timeValue\left\{{\redState}_{+1}\right\}
\end{bmatrix}   = \begin{bmatrix}
{\blueState} \\
{\redState}
\end{bmatrix} + 
\frac{1}{\numNodes}\times\begin{bmatrix}
\left(1-\blueState\right)\probBlueZeroToOne - \blueState\probBlueOneToZero\\
\left(1-\redState\right)\probRedZeroToOne - \redState\probRedOneToZero\\
\end{bmatrix},
\label{SIeq:FC_Expected_difference}
\end{equation}
where,
\begin{align*}
&\probBlueZeroToOne = \mathds{1}\left(\inAP(1-\redBirthProb) \left(2\blueContinuousTimeState - 1\right) - \outAP\redBirthProb\left(2\redContinuousTimeState-1\right) > 0\right), \hfill
&\probBlueOneToZero = 1- \probBlueZeroToOne\\
&\probRedZeroToOne = \mathds{1}\left(\inAP\redBirthProb\left(2\redContinuousTimeState - 1\right) - \outAP(1-\redBirthProb)\left(2\blueContinuousTimeState-1\right) > 0\right), \hfill
&\probRedOneToZero = 1- \probRedZeroToOne.
\end{align*}

Thus, the evolution of the state can be expressed as 
Eq.~\eqref{SIeq:Overview}. Note $g(\cdot)$ in Eq.~\eqref{SIeq:Overview} satisfies the linear growth condition since it is a piece-wise linear function taking values in $[-1,1]^2$. The set of discontinuities are defined by the states $\continuousTimeState(\continuouTimeValue) = [\blueContinuousTimeState(\continuouTimeValue), \redContinuousTimeState(\continuouTimeValue)]'$ that satisfy 
\begin{equation}
\inAP(1-\redBirthProb) \left(2\blueContinuousTimeState - 1\right) - \outAP\redBirthProb\left(2\redContinuousTimeState-1\right) = 0
\end{equation}
or
\begin{equation}
\inAP\redBirthProb\left(2\redContinuousTimeState - 1\right) - \outAP(1-\redBirthProb)\left(2\blueContinuousTimeState-1\right) = 0.
\end{equation}

Thus, Lemma~\ref{Lemma:Borkar2023} implies that the stochastic trajectory of the proposed model converges  to the solution set of the differential inclusion \begin{equation}
\label{SIeq:DI}\dot{\theta}(\continuouTimeValue) \in  h(\theta(\continuouTimeValue)) = [h^{\mathcal{B}}(\theta(\continuouTimeValue)), h^{\mathcal{R}}(\theta(\continuouTimeValue))]'\end{equation} 
where
\begin{align}
h^{\mathcal{B}}(\theta(\continuouTimeValue)) &= 
\begin{cases}
[-\blueContinuousTimeState(\continuouTimeValue), 1-\blueContinuousTimeState(\continuouTimeValue)] &\text{if } \inAP(1-\redBirthProb) \left(2\blueContinuousTimeState - 1\right) - \outAP\redBirthProb\left(2\redContinuousTimeState-1\right) = 0\\
\{g^{\mathcal{B}}(\theta)\}  &\text{otherwise,} 
\end{cases}\\
h^{\mathcal{R}}(\theta(\continuouTimeValue)) &= 
\begin{cases}
[-\redContinuousTimeState(\continuouTimeValue), 1-\redContinuousTimeState(\continuouTimeValue)] &\text{if } \inAP\redBirthProb\left(2\redContinuousTimeState - 1\right) - \outAP(1-\redBirthProb)\left(2\blueContinuousTimeState-1\right) = 0\\
\{g^{\mathcal{R}}(\theta)\}  &\text{otherwise,} 
\end{cases}
\end{align}
which is the Filippov solution set of the discontinuous differential equation Eq.~\ref{eq:FC_ODE}.

\subsection{Proof of Uniqueness}

To establish the uniqueness of the Filippov solution, we note that any solution to the Eq.~\ref{SIeq:DI} which approaches a point of discontinuity except $(0.5, 0.5)$ crosses the boundary and move to the next region. The only setting in which a trajectory approaches $(0.5, 0.5)$ is the Case~4 of Theorem~\ref{th:FC_Cases} with $\blueContinuousTimeState(0) = \redContinuousTimeState(0)$. Thus, according to \cite{cortes2008discontinuous}[Proposition~5], all trajectories except Case~4 with $\blueContinuousTimeState(0) = \redContinuousTimeState(0)$ are unique.

Uniqueness of trajectories can also be seen from state space plots for the four cases in 
Fig.~\ref{fig:Numerical_Results}
as well. Note that the only form of trajectory that approaches the boundary but does not cross in to the other region is the trajectory starting with $\blueContinuousTimeState(0) = \redContinuousTimeState(0)$ in Case~4. All other initial states therefore have unique trajectories.



\section{Dynamics of the Model on a Network with Communities}
\label{SI:SBM}

When the graph $\graph = (\nodeSet, \edgeSet)$ is a stochastic block model with in-group link probability $\homophily$ and out-group link probability of $1-\homophily$, the piece-wise interpolation of the discrete-time trajectory $\state = [\blueState, \redState]', \timeValue = 0, 1, 2, \dots$ can be approximated using the continuous-time trajectory $\continuousTimeState(\continuouTimeValue) = [\blueContinuousTimeState(\continuouTimeValue), \redContinuousTimeState(\continuouTimeValue)]', \continuouTimeValue\geq0$  of the following differential equation as the number of nodes in the graph $\numNodes$ is large:
\begin{align}
\label{eq:SBM_ODE}
   &\begin{bmatrix}
\DotBlueState\\
\DotRedState
\end{bmatrix}   = 
\begin{bmatrix}
\left(1-\blueContinuousTimeState\right)\probBlueZeroToOne - \blueContinuousTimeState\probBlueOneToZero\\
\left(1-\redContinuousTimeState\right)\probRedZeroToOne - \redContinuousTimeState\probRedOneToZero\\
\end{bmatrix},
\end{align}
where,
\begin{align*}
\probBlueZeroToOne &= \mathds{1}\left(\inAP\homophily(1-\redBirthProb) \left(2\blueContinuousTimeState - 1\right) - \outAP(1-\homophily)\redBirthProb\left(2\redContinuousTimeState-1\right) >\intertia  \right)\\
\probBlueOneToZero &= \mathds{1}\left(\inAP\homophily(1-\redBirthProb) \left(2\blueContinuousTimeState - 1\right) - \outAP(1-\homophily)\redBirthProb\left(2\redContinuousTimeState-1\right) <-\intertia  \right)\\
\probRedZeroToOne &= \mathds{1}\left(\inAP\homophily\redBirthProb\left(2\redContinuousTimeState - 1\right) - \outAP(1-\homophily)(1-\redBirthProb)\left(2\blueContinuousTimeState-1\right) >\intertia\right), \\
\probRedOneToZero &= \mathds{1}\left(\inAP\homophily\redBirthProb\left(2\redContinuousTimeState - 1\right) - \outAP(1-\homophily)(1-\redBirthProb)\left(2\blueContinuousTimeState-1\right) <-\intertia\right)
\end{align*}

\begin{figure}[hbt]
    \centering
    \includegraphics[width=\textwidth, trim=0.0in 0.1in 0.0in 0.0in, clip]{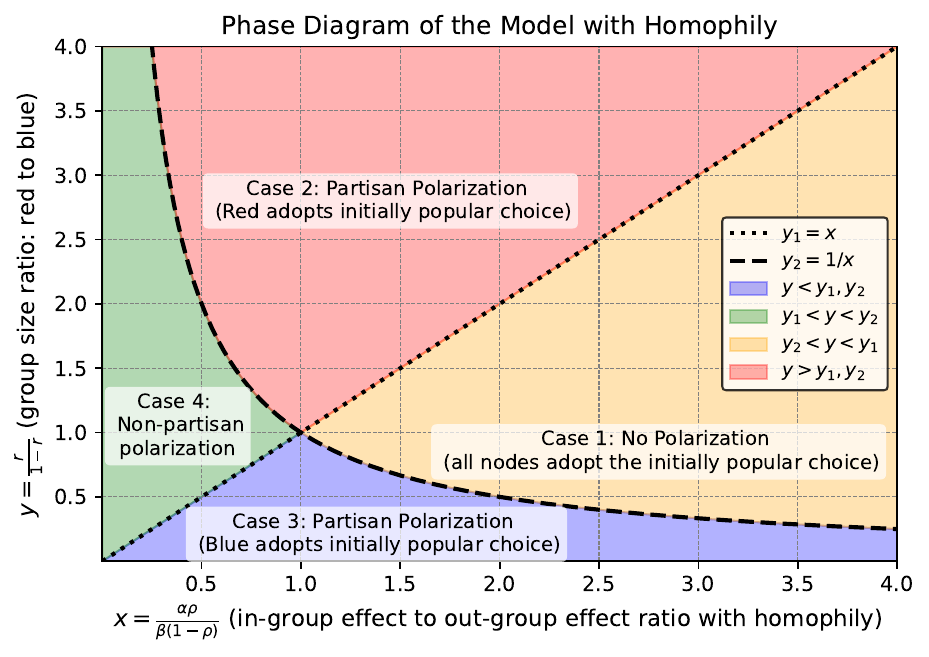}
    \caption{The four different regions of the model parameters (in-group conformity~$\inAP$, out-group dissent~$\outAP$, homophily~$\homophily$ and fraction of red-nodes $\redBirthProb$) that lead to different asymptotic behaviors in a stochastic block model type graph starting from an initial state where the distribution of choices is the same for both parties~i.e.,~i.e.,~$\blueContinuousTimeState(0) = \redContinuousTimeState(0)$). This figure is similar to the analogous figure for the fully connected graph~(Fig.~\ref{fig:Numerical_Results}) except that the in-group effect is amplified by $\homophily$~(probability observing each in-group member) and the out-group effect is amplified by $1-\homophily$~(probability of observing each out-group member).}
    \label{fig:SBM_Regions}
\end{figure}

Consequently, the analogous version of the Fig.~\ref{fig:Numerical_Results} for stochastic block models is shown in Fig.~\ref{fig:SBM_Regions}.

\section{Additional Results}

\begin{figure}
    \centering
    \vspace{-0.8in}
        \includegraphics[width=\textwidth, trim=0.0in 0.12in 0.0in 0.0in, clip]{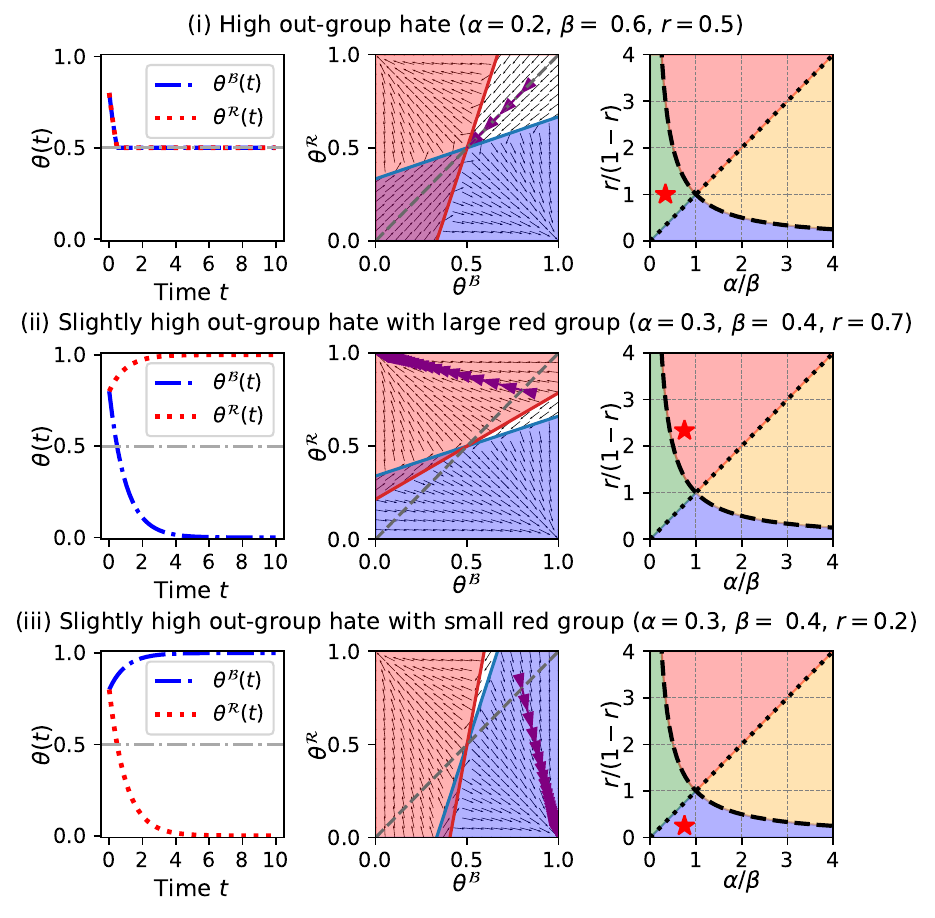}
    \caption{Example trajectories  of the state 
    when the out-group hate~$\outAP$ is larger than in-group love $\inAP$. The trajectories $[\blueContinuousTimeState(\continuouTimeValue), \redContinuousTimeState(\continuouTimeValue)]$ over time~(left column) and in the state space~(middle column) show the evolution of $\continuousTimeState(\continuouTimeValue) = [\blueContinuousTimeState(\continuouTimeValue), \redContinuousTimeState(\continuouTimeValue)]$.  The blue and red colors in middle column indicate regions  where $\blueContinuousTimeState(\continuouTimeValue), \redContinuousTimeState(\continuouTimeValue)$ increase~(i.e.,~regions where $\probBlueZeroToOne = 1$ and $\probRedZeroToOne=1$ according to Eq.~\ref{eq:FC_ODE}). The black arrows in state space plots~(middle column) indicate the path of the differential equation Eq.~\ref{eq:FC_ODE}. The yellow arrows corresponds to the time domain trajectory~(in left column). The figure shows how either 
    uniform~(row~i) or party-line polarization~(row~ii and row-iii) can emerge when people are driven largely by their opposition to the out-group than their adherence to the in-group. Further, uniform polarization that emerges in the presence of very high out-group hate is unstable since some black arrows point away from $[0.5,0.5]$ as seen from the state space plots~(middle column) of row-i. In this case, small deviations from non-partisan polarization can lead to partisan polarization .}
    \label{fig:LargeBeta}
\end{figure}

\begin{figure}
    \centering
    \vspace{-0.8in}
        \includegraphics[width=\textwidth, trim=0.0in 0.12in 0.0in 0.0in, clip]{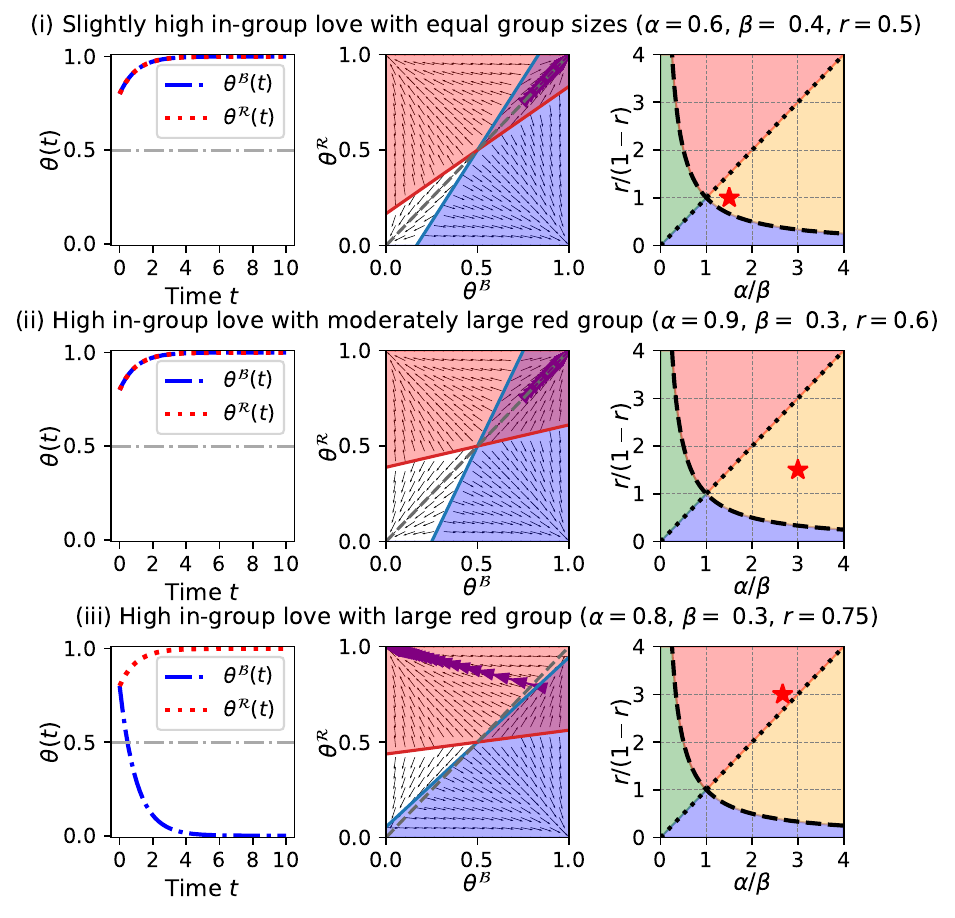}
    \caption{Example trajectories of the state 
    when out-group hate~$\outAP$ is less than in-group love $\inAP$. The trajectories over time~(left column) and in the state space~(middle column) show the evolution of $\continuousTimeState(\continuouTimeValue) = [\blueContinuousTimeState(\continuouTimeValue), \redContinuousTimeState(\continuouTimeValue)]$. The blue and red regions in middle column indicate regions where $\blueContinuousTimeState(\continuouTimeValue), \redContinuousTimeState(\continuouTimeValue)$ increase~(i.e.,~areas where $\probBlueZeroToOne = 1$ and $\probRedZeroToOne=1$ according to Eq.~\ref{eq:FC_ODE}). The black arrows in the middle column indicate the path of the differential equation Eq.~\ref{eq:FC_ODE}. The yellow arrows  correspond to the time domain trajectory~(left column). The figure shows how larger in-group love is necessary but not sufficient for the emergence of consensus. In particular, when the disparity between the sizes of the two groups is not too large compared to the disparity between $\inAP$ and $\outAP$, consensus emerges.}
    \label{fig:LargeAlpha}
\end{figure}

\begin{figure}
    \centering
        \includegraphics[width=1\textwidth, trim=0.0in 0.12in 0.0in 0.0in, clip]{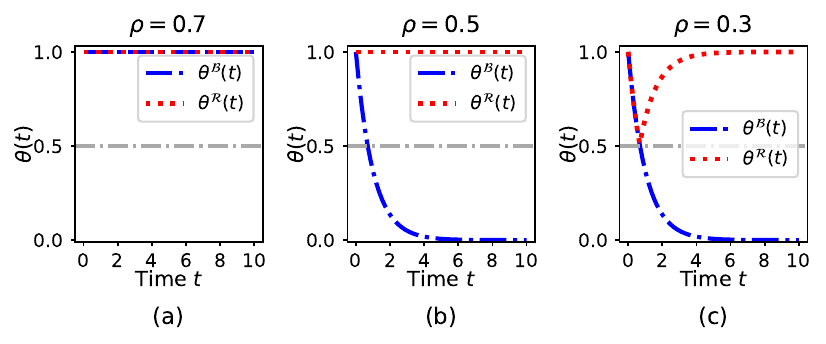}
    \caption{An illustration of how decreasing homophily can cause a party-line polarization from an initial state of global consensus. Figures correspond to $\inAP = 0.8, \outAP = 0.7$~(larger in-group favoritism compared to out-group animosity) and $\redBirthProb = 0.65$~(a majority red group). Decreasing $\homophily$ from 0.7~(homophily) to 0.5 pushes the social network from Case-1~(consensus) to Case-3~(party-line polarization) in Fig.~\ref{fig:SBM_Regions}. Further decreasing $\homophily$ to 0.3 pushes the network to Case-4 which corresponds to an unstable state, where a small deviation leads to party-line polarization.}
    \label{fig:BreakingUnity}
\end{figure}

\begin{figure}
    \centering
        \includegraphics[width=1\textwidth, trim=0.0in 0.1in 0.0in 0.0in, clip]{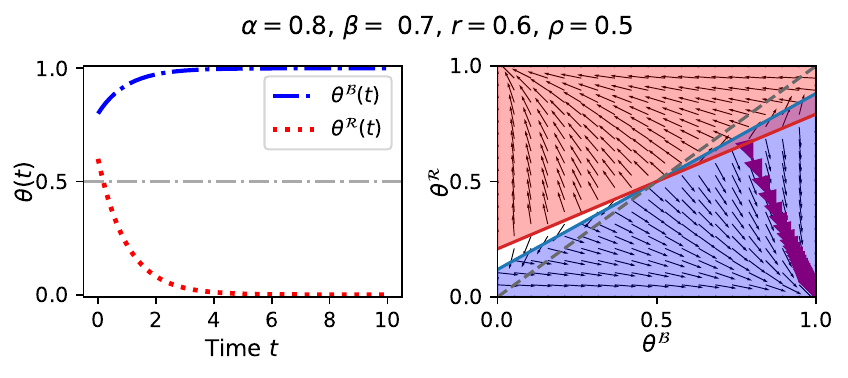}
    \caption{An illustration of a case where the two groups start with different popularity levels of the choices within them~i.e.,~$\blueContinuousTimeState(0) \neq \redContinuousTimeState(0)$, and the majority group adopts the choice that was initially less popular within it. The choice-1 is initially more popular within both groups with $\blueContinuousTimeState(0) =0.8, \redContinuousTimeState(0) = 0.6$. However, the majority red-group eventually adopts the choice that was initially less popular~(i.e.,~choice-0 which had a 40\% popularity) within it.}
    \label{fig:DifferentInitialStates_majority}
\end{figure}
\end{document}